\documentclass[english,12pt]{article}
          \usepackage{babel}
          \usepackage[T1]{fontenc}
          \usepackage[latin2]{inputenc}
          \usepackage{pslatex}
\begin{document}

\title{\bf Superhumps in V1141 Aquilae}
\author{Arkadiusz ~ O~l~e~c~h}
\date{Nicolaus Copernicus Astronomical Center, \\ 
Polish Academy of Sciences, \\
ul.~Bartycka~18, 00-716~Warszawa, Poland\\ 
{\tt e-mail: olech@camk.edu.pl}}

\maketitle

\begin{abstract}
The results of the CCD observations of the 2002 superoutburst of V1141
Aql are described. We have detected clear superhumps characterized by
the period of 0.05930(5) days. There was another, much weaker, 
modulation in the light curve of V1141 Aql. Its amplitude is equal to
$0.012\pm0.002$ mag and its period to 0.03923(8) days. We have also
discovered quasi periodic oscillations (QPOs) with a mean period equal
to 130 s and amplitude of $0.025\pm0.004$ mag.

\end{abstract}

\section{Introduction}

V1141 Aql was suspected for variability by Kukarkin {\it et al.} (1951)
and confirmed as a variable star by Kukarkin {\it et al.} (1968). It was
later classified as a dwarf nova by Vogt and Bateson (1982). In
subsequent documentation, V1141 Aql is only referred to as a dwarf nova
(Bruch {\it et al.} 1992, Downes {\it et al.} 1997).

We learnt about the continuing outburst of V1141 Aql from the VSNET
e-mail \footnote{[vsnet-outburst-4198] 
http://vsnet.kusastro.kyoto-u.ac.jp/vsnet/Mail/vsnet-outburst/msg04198.html}
reporting the observation of Maciej Reszelski and indicating that the star
reached a magnitude of 15.2 at 2002 July 8.927 UT. Previous brightenings
of V1141 Aql were reported on 2001 July and 2000 September (Reszelski
2002, private communication).

\section{Observations}

Observations of V1141 Aql reported in the present paper were obtained
over five nights during the period from 2002 July 9 to 14 at the
Ostrowik station of the Warsaw University Observatory.

The 60-cm Cassegrain telescope equipped with a Tektronics TK512CB back
illuminated CCD camera was used. The scale of the camera was 0.76"/pixel
providing a $6.5'\times 6.5'$ field of view. The full description of the
telescope and camera was given by Udalski and Pych (1992).

We monitored the star in ``white light''. This was due to the lack of
an autoguiding system, not yet implemented after recent telescope
renovation. Thus we did not use any filter because we wanted to have
the shortest possible exposure times and avoid elongation of the star
shapes.

The exposure times were from 60 to 150 seconds depending on the
atmospheric conditions, seeing and the brightness of the object.

A full journal of our CCD observations of V1141 Aql is given in Table
1. In total, we monitored the star during 13.71 hours and obtained 436
exposures.

\begin{table}[h]
\caption{\sc Journal of the CCD observations of V1141 Aql}
\vspace{0.1cm}
\begin{center}
\begin{tabular}{|l|c|c|c|}
\hline
\hline
Date & Time of start & Length of & Number of \\
     & 2452000. + & run [hr] & frames \\
\hline
\hline
2002 Jul 09/10 & 465.4379 & 2.38 & 83 \\
2002 Jul 10/11 & 466.4358 & 2.41 & 83 \\
2002 Jul 12/13 & 468.4532 & 2.34 & 84 \\
2002 Jul 13/14 & 469.4181 & 3.09 & 91 \\
2002 Jul 14/15 & 470.3911 & 3.49 & 95 \\
\hline
Total          &   --   & 13.71 & 436 \\ 
\hline
\hline
\end{tabular}
\end{center}
\end{table}

\subsection{Data Reduction}

All the data reductions were performed using a standard procedure
based on the IRAF \footnote{IRAF is distributed by the National Optical
Astronomy Observatory, which is operated by the Association of
Universities for Research in Astronomy, Inc., under a cooperative
agreement with the National Science Foundation.} package.

The profile photometry has been derived using the DAOphotII package
(Stetson 1987).

Relative unfiltered magnitudes of V1141 Aql were determined as the
difference between the magnitude of the variable and the magnitude
corresponding to the sum of intensities of four comparison stars. These
comparison stars are marked in the chart displayed in Fig. 1. The
accuracy of our measurements varied between 0.008 and 0.072 mag
depending on the atmospheric conditions and brightness of the variable.
The median value of the photometry errors was 0.015 mag.

\subsection{Light curves}

Fig. 2 presents a general photometric behavior of V1141 Aql as observed
in July 2002. The brightness of the star is at the same level during the
first two nights indicating that the variable was caught at the beginning
of its superoutburst. During the subsequent nights one can see a systematic
decrease of the brightness with a rate of 0.15 mag per day.

Fig. 3 presents the individual light curves of the star for our five
nights. In all observations the short-term modulation (i.e. superhumps)
is clearly visible, confirming that V1141 Aql is a member of the SU UMa
group of dwarf novae.

\clearpage
~

\vspace{10cm}

\includegraphics{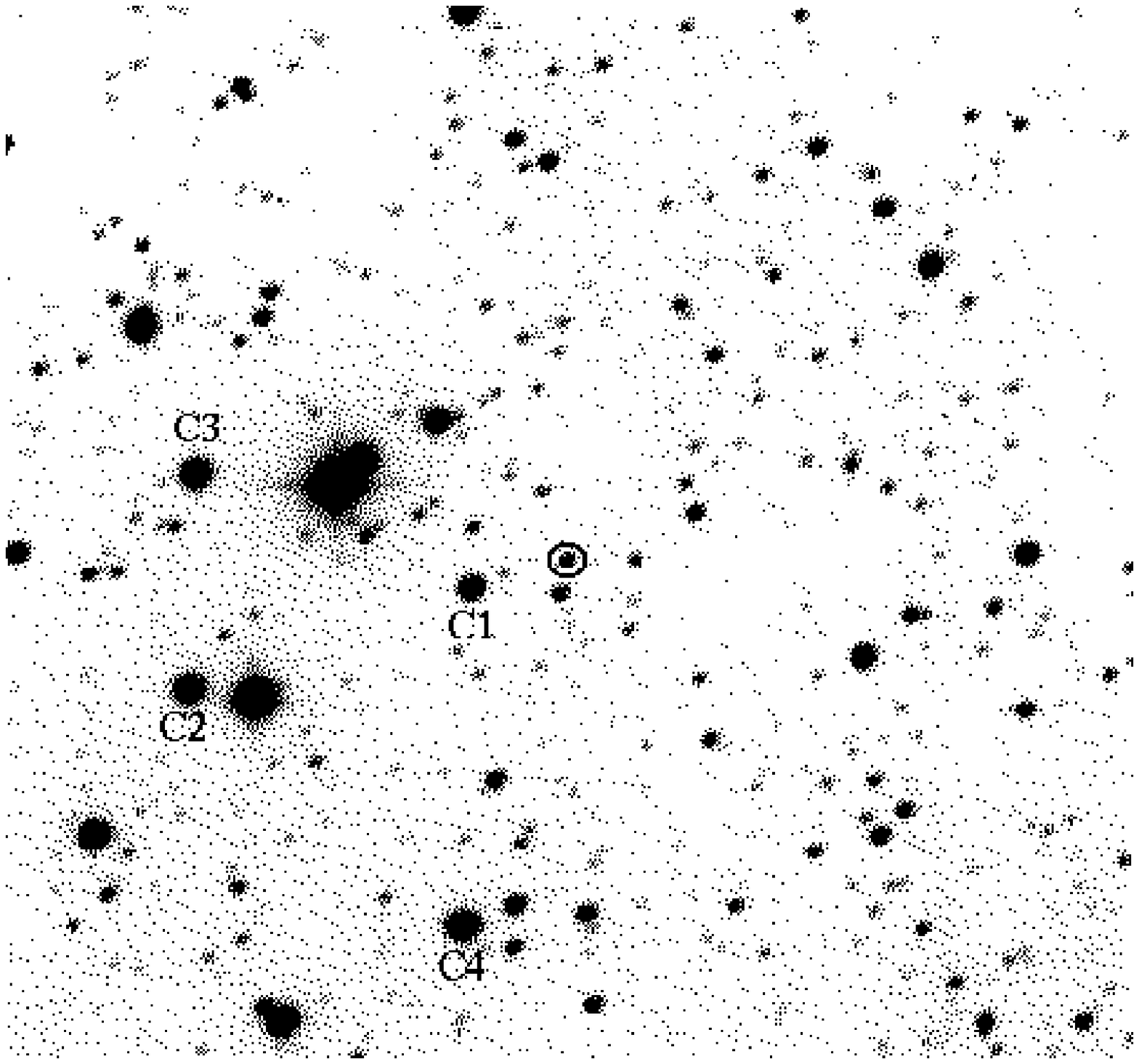}

\begin{figure}[h]
\caption{\sf Finding chart for V1141 Aql covering a region of $6.5 \times
6.5$ arcminutes. The variable is marked by an open circle. The positions
of the comparison stars are also shown.}
\end{figure}

\vspace{8.5cm}

\includegraphics{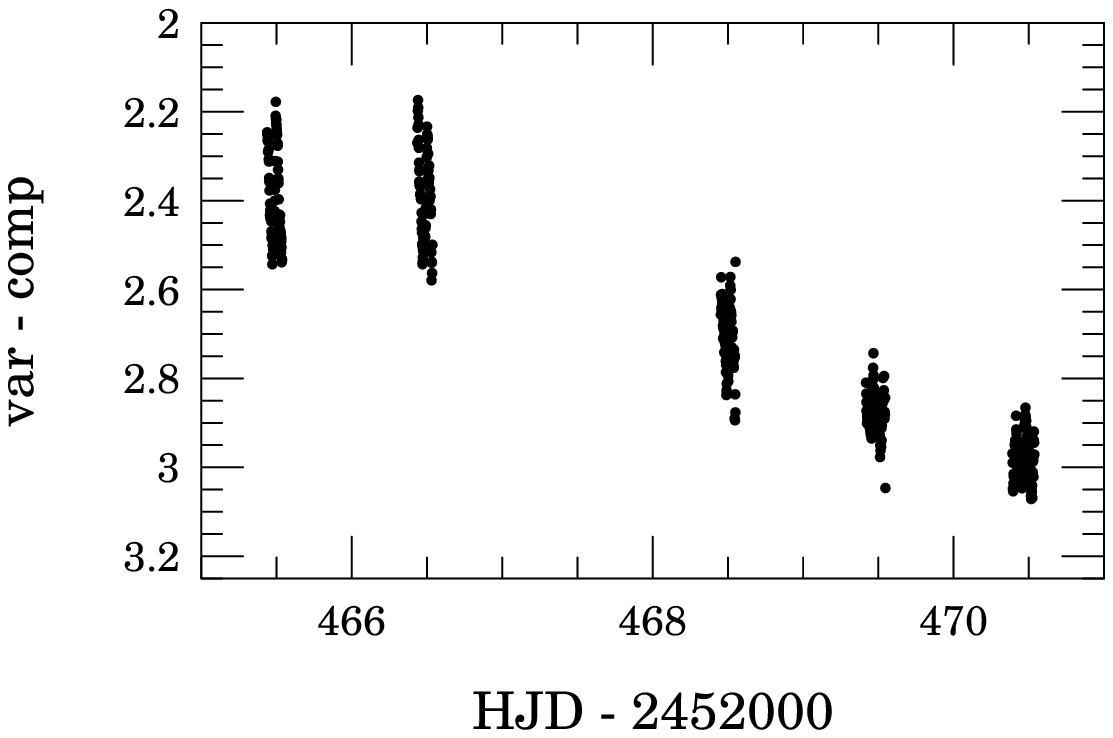}

\begin{figure}[h]
\caption{\sf The general photometric behavior of V1141 Aql during
its 2002 superoutburst}
\end{figure}
\clearpage
~

\vspace{19.9cm}

\includegraphics{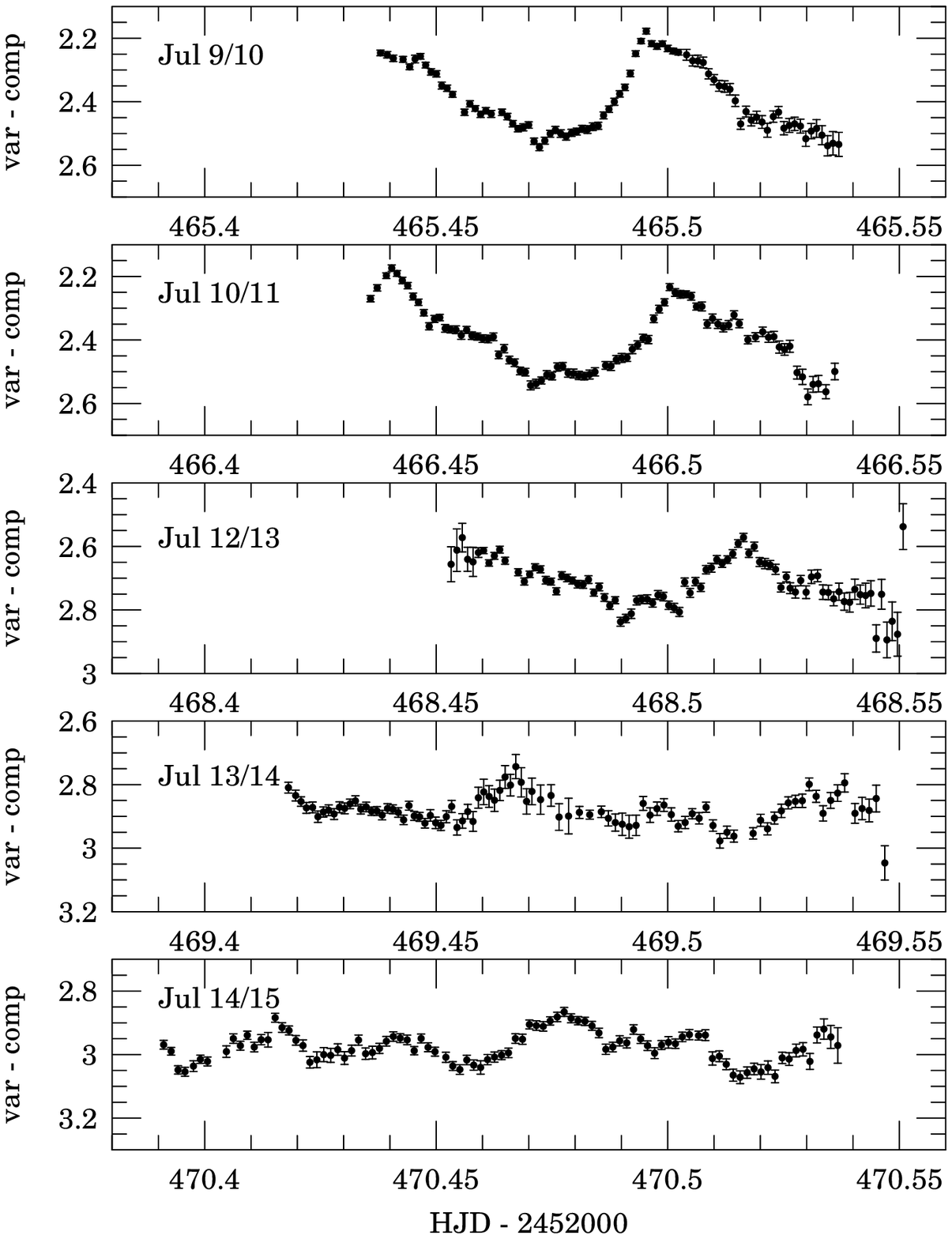}

\begin{figure}[h]
\caption{\sf The light curves of V1141 Aql observed during five
nights in July 2002}
\end{figure}

\section{Period of the Superhumps}

\subsection{Power spectrum}

From the light curves we removed the long term decrease and then
analyzed them using {\sc anova} statistics (Schwarzenberg-Czerny 1996).
The resulting periodogram is shown in Fig. 4. The most prominent peak is
detected at a frequency of $f=16.878\pm0.020$ c/d, which corresponds to
the period $P_{sh}=0.05928(7)$ days ($85.36\pm0.10$ min). The first
harmonic of this frequency is also marginaly visible around 32-34 c/d.

\vspace{5.2cm}

\includegraphics{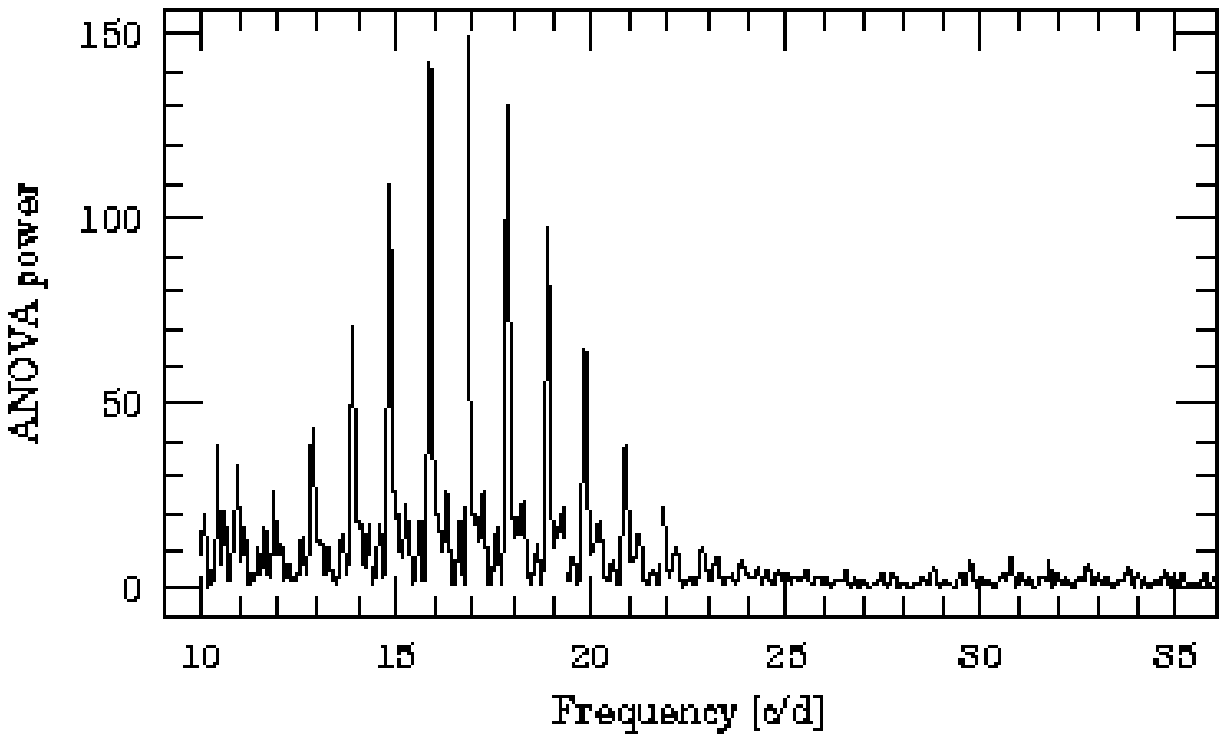}

\begin{figure}[h]
\caption{\sf {\sc anova} power spectrum of the light curve of V1141 Aql.}
\end{figure}
\smallskip

The light curve from each night was separately fitted with the sum of
six harmonics characterized by a period of 0.05928 days. In the next
step, we have prewhitened our light curves removing the variability with
$P_{sh}$. The {\sc anova} power spectrum of the resulting light curve is
shown in Fig. 5. One can clearly see a distinct peak at $f_2=25.49\pm0.05$
c/d, which corresponds to a period of $P_2=0.03923(8)$ days ($56.5\pm0.1$
min). The inset shows the prewhitened light curve of V1141 Aql phased
over this period. The amplitude of this modulation is equal to
$0.012\pm0.002$ mag. 

\vspace{6.6cm}

\includegraphics{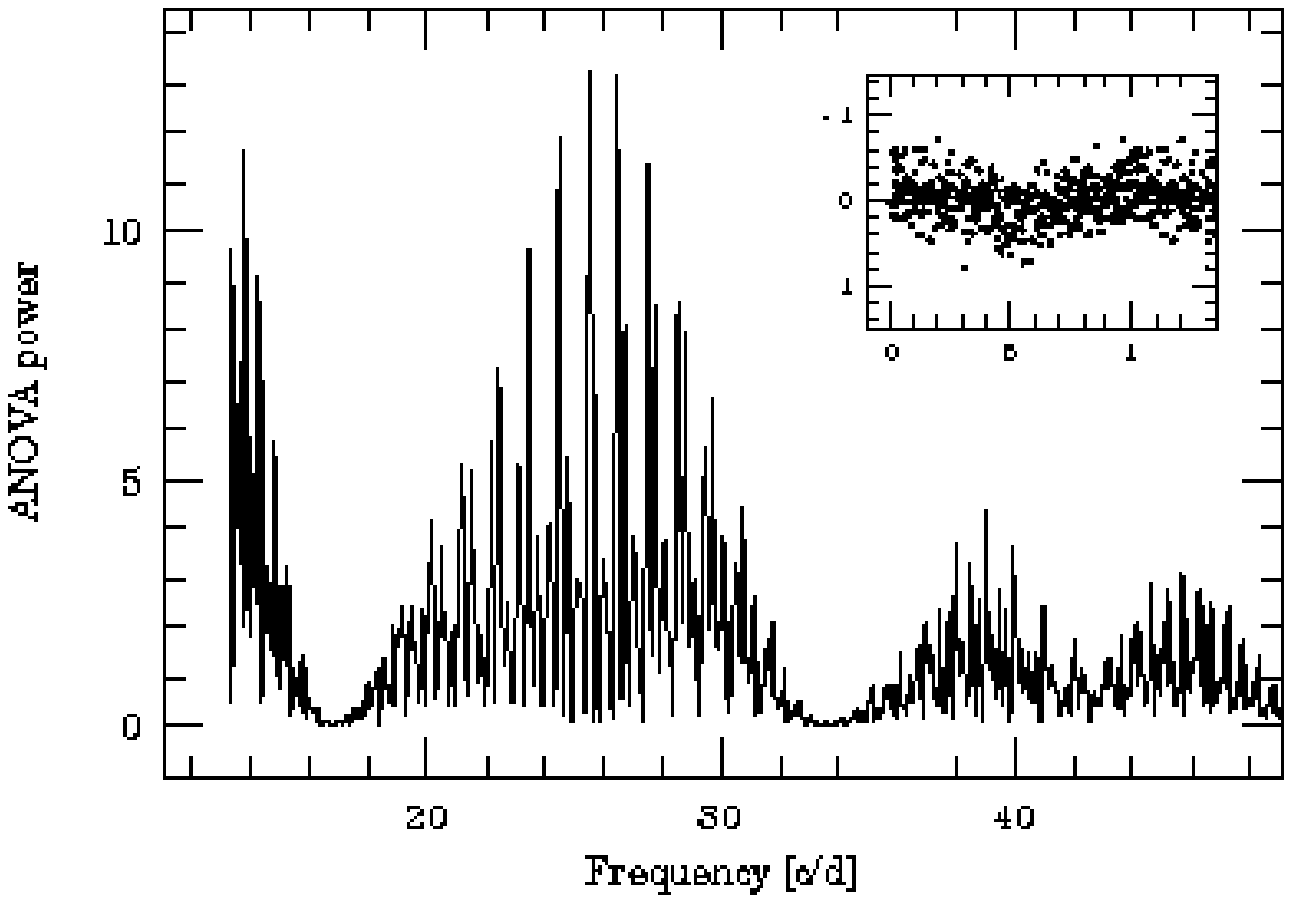}

\begin{figure}[h]
\caption{\sf {\sc anova} power spectrum of the light curve 
of V1141 Aql prewhitened with $P_{sh}$. Inset shows the 
light curve phased with a period of 0.03923 
days corresponding to the highest peak.}
\end{figure}
\clearpage

To verify the accuracy of this period we performed a separate frequency
analysis of each night. The clear peaks at a frequency of approximately
25.5 c/d were present in all resulting periodograms indicating that the
new period is most probably a real feature.

However one can clearly see that in the periodogram shown in Fig. 5 the
peak with frequency 26.49 c/d has almost the same power as peak near
25.5 c/d. Thus we can not exclude that the frequency 26.49 c/d is the
real one and the true period of 0.01 mag modulations is equal to 0.03775
days.

\subsection{O -- C Diagrams}

One of the most popular methods for determining the period of the light
modulation and its time derivative is the formulation of the O -- C diagram
for the times of extrema. In our case the maxima were clearer than the
minima and thus we decided to use them in our analysis. 

\begin{table}[h]
\caption{\sc Times of maxima observed in the light curve
of V1141 Aql during its 2002 superoutburst}
\vspace{0.1cm}
\begin{center}
\begin{tabular}{|l|c|r|}
\hline
\hline
$E$ & $HJD_{max}$ & O -- C \\
    &             & [cycles] \\
\hline
0  & 2452465.4960 & $ 0.0337$\\
16 & 2452466.4410 & $-0.0357$\\
17 & 2452466.5020 & $-0.0074$\\
51 & 2452468.5165 & $-0.0475$\\
67 & 2452469.4670 & $-0.0243$\\
68 & 2452469.5330 & $ 0.0883$\\
83 & 2452470.4150 & $-0.0432$\\
84 & 2452470.4780 & $ 0.0189$\\
\hline
\hline
\end{tabular}
\end{center}
\end{table}
\bigskip

In total, we determined eight maxima peaks listed in Table
2 together with their cycle numbers $E$. The least squares linear fit to
the data from Table 2 gives the following ephemeris:

\begin{equation}
HJD_{max} =  2452465.494(3) + 0.05932(6) \cdot E
\end{equation}

\noindent indicating that the value of the superhump period $P_{sh}$ is
equal to 0.05932(6) days ($85.42\pm0.09$ min). This is in excellent
agreement with the value obtained in the previous paragraph from the
power spectrum analysis.

Table 2 also shows O -- C data computed using our times of maxima and
ephemeris (1). They are also shown in Fig. 6. Larger O -- C values
observed at the beginning and the end of our observation period may
suggest that the period of the superhumps is increasing. With a data
span interval of only six days the assertion that there has been a
change in the period is only hypothetical. However the formal fit of the
second order polynomial results in the following relation:

\begin{equation}
HJD_{max} =  2452465.495(3) + 0.05925(15) \cdot E + 1.0(1.7)\times 10^{-6} E^2
\end{equation}

\noindent indicating that the period was likely increasing with a rate of
$\dot P \approx 3 \times 10^{-5}$.

This is not typical behavior among SU UMa stars where we often
observe decreasing periods of superhumps. 

Finally, by combining our two determinations of $P_{sh}$, we derive its
value as 0.05930(5) days, which corresponds to $85.39 \pm 0.07$ min.

\vspace{7.6cm}

\includegraphics{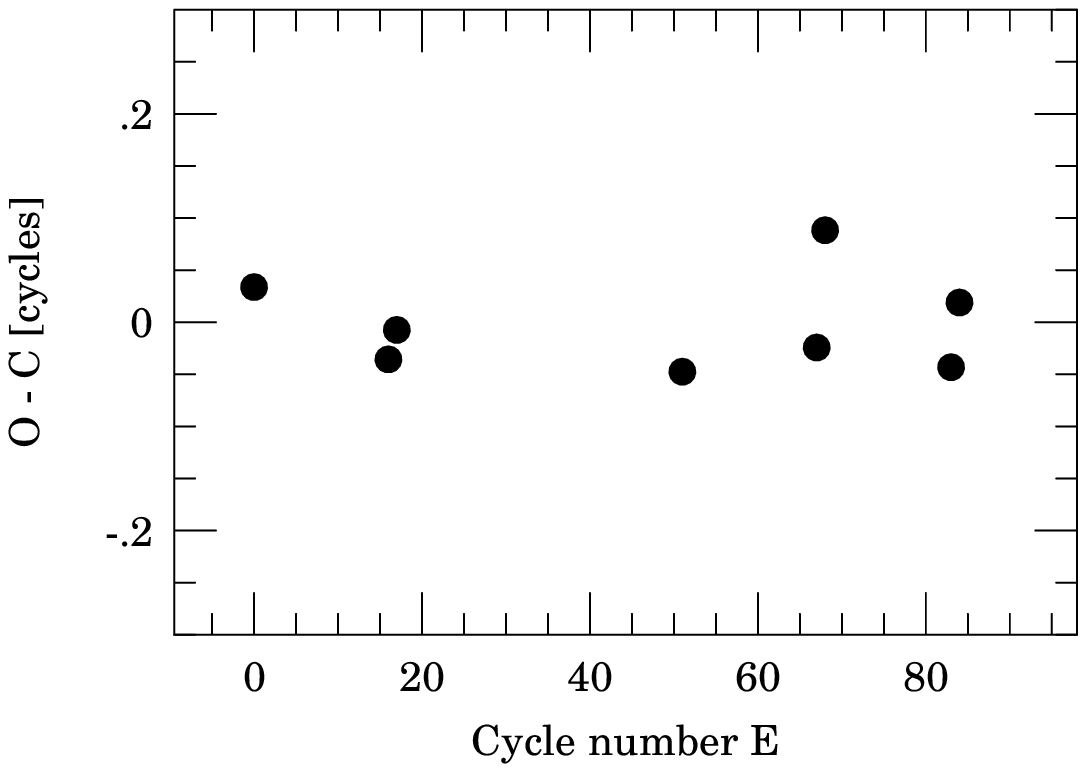}

\begin{figure}[h]
\caption{\sf O -- C diagram for 8 times of superhump maxima of V1141 Aql.}
\end{figure}

\section{Quasi Periodic Oscillations}

Quasi periodic oscillations (QPOs) are low amplitude and short period
luminosity modulations observed in the light curves of cataclysmic
variables. They were first discussed by Patterson {\it et al.} (1977).
The periods of these modulations are in the range of $\sim$50-1000 s and
their amplitudes are usually only a few times of 0.01 mag. However
occasionally larger amplitudes are observed. For example, in 1974 QPOs
in VW Hyi had an amplitude reaching 0.12 mag (Warner \& Brickhill 1978).
The most prominent modulations with an amplitude of 0.2 mag were
observed by Kato {\it et al.} (1992) during the 1992 superoutburst of SW
UMa. During two other superoutbursts of this star in 1986 and 1996 QPOs
were also detected but with significantly smaller amplitudes (Robinson
{\it et al.} 1987, Nogami {\it et al.} 1998).

We decided to search for QPOs in the light curve of V1141 Aql. For this
purpose we prewhitened our light curve using two periods described in
section 3.1. The variations with $P_{sh}$ and its six harmonics were
subtracted from the light curve of the star for each night separately.
From the resulting light curve we removed sinusoid characterized by 
the period 0.03923 days and an amplitude of 0.012 mag. For such a
prewhitened light curve we computed the {\sc anova} power spectrum for
frequencies in the range of 0 - 900 c/d. This is shown in Fig. 7.

There is a clear group of distinct peaks around frequency 670 c/d in
Fig. 7. The subharmonic of this frequency at 335 c/d is also visible.
The highest peak in this power spectrum has a frequency of 663.8 c/d
which corresponds to the period of 130 s. The question is whether this
frequency is real in view of the fact that the typical exposure
time used in our observations was 90 s, i.e. almost 70\% of the detected
period.

The mean amplitude of these 130 s QPOs in V1141 Aql is equal to
$0.0103\pm0.0015$ mag. Due to the fact that during our observations we
averaged the brightness of the stars over 70\% of the period, the
true amplitude of the suspected QPOs is higher. This can be easily computed
from the following expression connecting the real amplitude $A_r$ with
observed amplitude $A_o$, the period of the QPOs $P_{QPO}$ and exposure
time $T_{exp}$:

\begin{equation}
A_r = {A_o { {\pi T_{exp}} \over {P_{QPO}}} \cdot {\sin ^{-1}({ {\pi T_{exp}} \over{P_{QPO}} })} }
\end{equation}

\noindent Thus finally the real amplitude of the QPOs in V1141 Aql is
$0.025\pm0.004$ mag.

\vspace{7cm}

\includegraphics{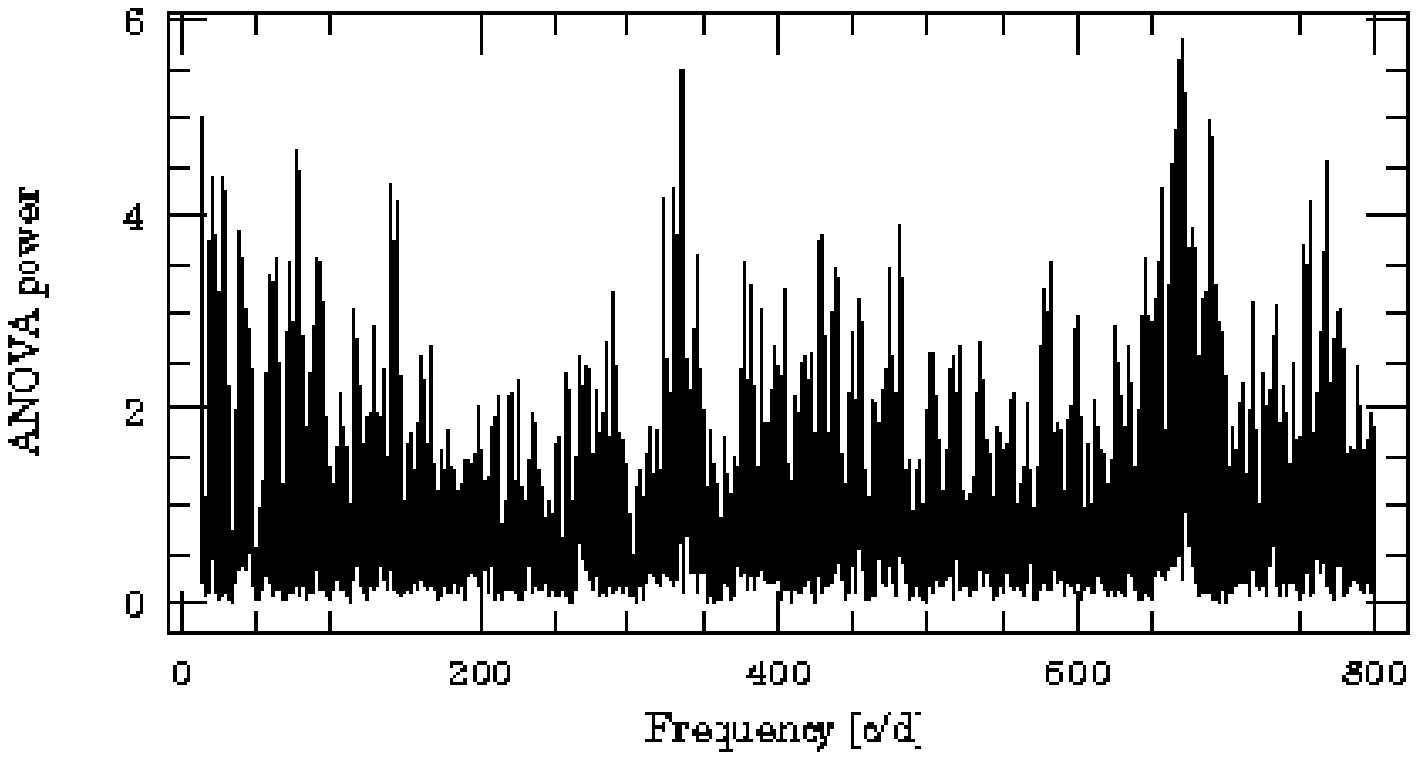}

\begin{figure}[h]
\caption{\sf {\sc anova} power spectrum of the prewhitened light curve
of V1141 Aql showing the clear group of peaks around frequency 670 c/d }
\end{figure}

\vspace{6.8cm}

\includegraphics{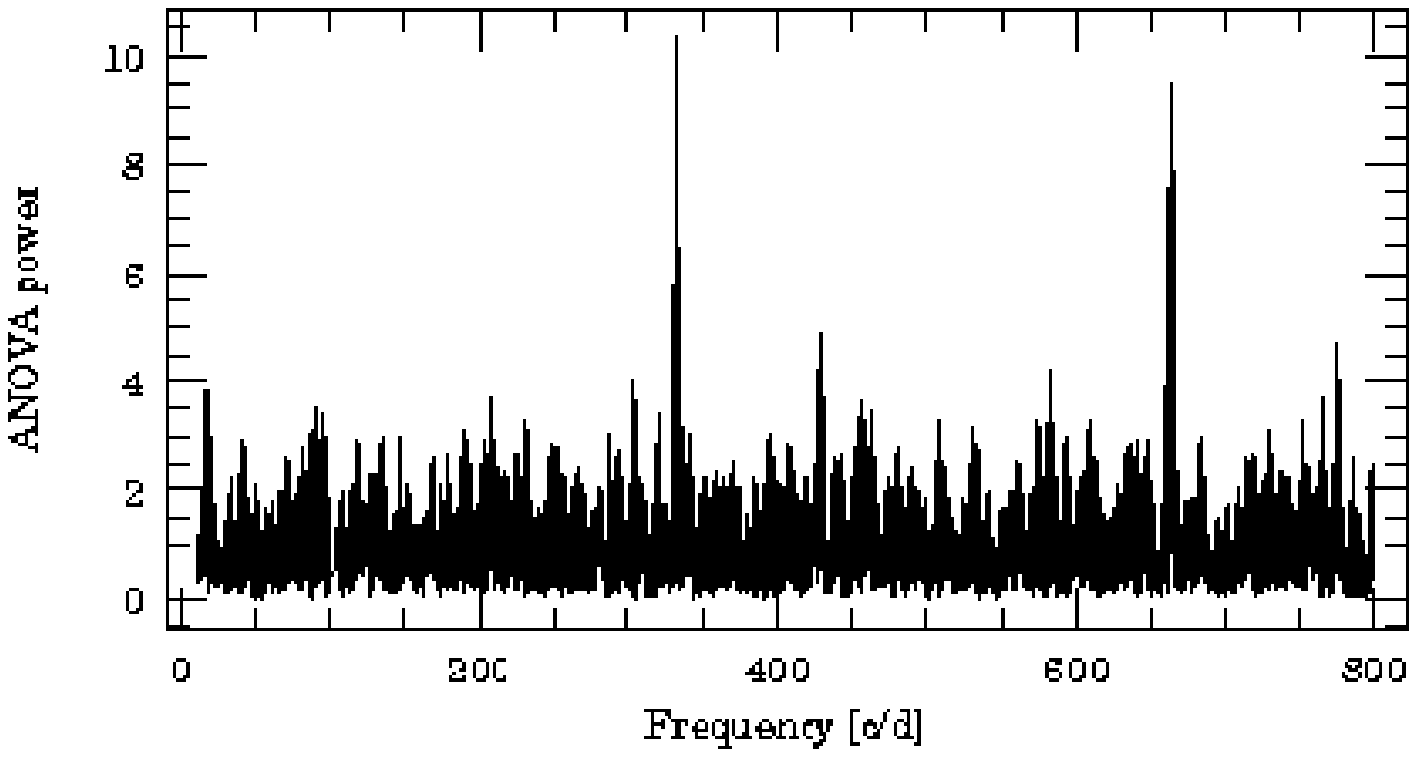}

\begin{figure}[h]
\caption{\sf {\sc anova} power spectrum of the simulated light curve
of V1141 Aql}
\end{figure}

To verify the reality of the QPOs in the light curve of V1141 Aql we
performed a simple test. After prewhitenning the raw light curve of the
star with all three detected periods there is no any periodic signal
left. The observed noise is characterized be the Gaussian function with
$\sigma$ equal to 0.024 mag. Thus we constructed an artificial light
curve using a sinusoid with an amplitude of 0.025 mag and a period of
130 s. For each of our observations we computed the mean brightness of
this artificial curve averaging it over $\pm45$ s around the time of
observation. Then the noise described by the Gaussian function with
$\sigma=0.024$ mag was added. For such a constructed light curve we
computed the {\sc anova} power spectrum, which is shown in Fig. 8. Both
power spectra shown in Figs. 7 and 8 are similar, suggesting that 130 s
QPOs in the light curve of V1141 Aql are real phenomena. Of course,
there is more noise in the power spectrum of the real light curve and it
may be caused by a high order of linear combinations of frequencies
described in section 3.1 which were not removed during prewhitenning.
The source of the additional noise is certainly related to the
quasi-periodic behavior of the detected oscillations. In our simulated
light curve we inputed only the sinusoidal variations with the constant
value of the period.

Finally, we conclude that the light curve of 2002 superoutbursts of V1141
Aql contained QPOs with a mean period of 130 s and an amplitude of
$0.025\pm0.004$ mag.

\section{Discussion}

We have presented the results of the observations of the 2002
brightening of V1141 Aql. Detection of the clear superhumps during this
event directly proves that V1141 Aql belongs to the group of SU UMa
variables.

The value of the superhump period was equal to 0.05930(5) days which
corresponds to $85.39 \pm 0.07$ min. Our data seem to indicate that
the superhump period was increasing at a rate of $\dot P \approx 3
\times 10^{-5}$. It seems to agree with the pattern observed in other
SU UMa systems, where long-period systems show a decrease of the
superhump periods, and short-period systems or infrequently outbursting
SU UMa variables tend to show an increase of the superhump period (see
Kato {\it et al.} 2001a for discussion).

Previous brightenings of V1141 Aql were observed in July 2001 and
September 2000 (Reszelski 2002, private communication). The star then
reached a magnitude of approximately 15.5. This may indicate that these
brightenings were also superoutbursts and the supercycle in V1141 Aql
lasts slightly less than one year.

The interesting thing is lack of ordinary outbursts which may be real
phenomenon suggesting a resemblance of V1141 Aql to the group of
"tremendous outburst amplitude dwarf novae" or "TOADs" (Howell {\it et
al.} 1995). On the other hand, the lack of the ordinary outburst may be
an observational effect as well. V1141 Aql, even in the superoutburst,
is faint. Ordinary outbursts being 1-2 mag fainter than superoutbursts
could be below the limiting magnitude reached by astronomy amateurs.

Other properties of V1141 Aql are also similar to those shown by TOADs.
The star seems to be a twin of SW UMa - a well known member of TOADs. SW
UMa was intensively studied during its 1996 superoutburst. According to
Semeniuk {\it et al.} (1997) the superhump period of SW UMa is only
slightly shorter that the period of V1141 Aql. Thus both stars have
similar superhump periods, similar supercycle lengths and are
characterized by the presence of interpulses (secondary humps) visible
on the light curves in the second part of the superoutburst. Both stars
also show clear QPOs in their superoutburst light curves and for both we
do not observe ordinary outbursts. SW UMa was the first variable of SU
UMa type with an increasing period of superhumps (Semeniuk {\it et al.}
1997). Such behavior was observed only in the other short orbital
period SU UMa stars such as V485 Cen (Olech 1997), AL Com (Nogami
{\it et al.} 1997), V1028 Cyg (Baba {\it et al.} 2000) and HV Vir (Kato
{\it et al.} 1998) and WX Cet (Kato {\it et al.} 2001b).

A comparison of the properties of V1141 Aql and SW UMa is shown in Table
3.

\begin{table}[h]
\caption{\sc Comparison of the properties of V1141 Aql and SW UMa}
\vspace{0.1cm}
\begin{center}
\begin{tabular}{|l|c|c|}
\hline
\hline
 & V1141 Aql & SW UMa \\
\hline
Supercycle & 350 days & 400 days \\
Amplitude of superoutburst & 5 mag. & 7 mag. \\
Superhump period & 85.4 min. & 83.8 min. \\
Orbital period & ? & 81.8 min. \\
$\dot P_{sh}$ & $+3 \times 10^{-5}$ & $+8.9 \times 10^{-5}$ \\
Interpulses & yes & yes \\
QPOs & yes & yes \\
Ordinary outbursts & no? & no\\
\hline
\hline
\end{tabular}
\end{center}
\end{table}

Of course, we can not include V1141 Aql into the TOADs because according
to Howell {\it et al.} (1995) the main property of this group is the
amplitude of the superoutbursts higher than 6 mag. In the case of V1141 Aql
this amplitude is around 5 mag, suggesting that this star may be a
link between ordinary SU UMa stars and TOADs.

In addition to the clearly visible modulation connected with the
superhumps we have also detected the weak signal characterized by an
amplitude around 0.01 mag and a period of 0.03923(8) days ($56.5\pm0.1$
min). The nature of this periodicity is unknown. This cannot be the
orbital period of the binary system because, according to the Schoembs
\& Stolz (1984) relation, it should be around 84 min. It could be a
signature of the white dwarf rotation period, which in nonmagnetized
systems such as SU UMa stars, is not synchronized with the orbital
period.

We have also discovered quasi periodic oscillations (QPOs) with a mean
period equal to 130 s and an amplitude of $0.025\pm0.004$ mag.

\bigskip

\noindent {\bf Acknowledgments.} We would like to thank to Prof. J\'ozef
Smak and Chris O'Connor for reading and commenting on the manuscript.

\end{document}